\documentclass[a4paper,superscriptaddress,prb,showkeys,twocolumn]{revtex4}

\usepackage{graphicx} 
\newcommand{\sglcolfigure}[1]%
{\includegraphics[width=8.2cm,keepaspectratio]{#1}}
\newcommand{\dblcolfigure}[1]%
{\includegraphics[width=16.8cm,keepaspectratio]{#1}}

\begin{document}

\title{k-space Imaging of the Eigenmodes of Sharp Gold Tapers for Scanning Near-Field Optical Microscopy}
\author{Martin Esmann}
\author{Simon F. Becker}
\author{Bernard B. da Cunha}
\author{Jens H. Brauer}
\author{Ralf Vogelgesang}
\author{Petra Gro\ss}
\author{Christoph Lienau}
\thanks{Corresponding author}
\email{christoph.lienau@uni-oldenburg.de}
\affiliation{Institut f\"ur Physik, Carl von Ossietzky Universit\"at, 26129 Oldenburg, Germany}
\affiliation{Center of Interface Science, Carl von Ossietzky Universit\"at, 26129 Oldenburg, Germany}

\begin{abstract}
We investigate the radiation patterns of sharp conical gold tapers, designed as adiabatic nanofocusing probes for scanning near-field optical microscopy (SNOM). Field calculations show that only the lowest order eigenmode of such a taper can reach the very apex and thus induce the generation of strongly enhanced near-field signals. Higher order modes are coupled into the far field at finite distances from the apex. Here, we demonstrate experimentally how to distinguish and separate between the lowest and higher order eigenmodes of such a metallic taper by filtering in the spatial frequency domain. Our approach has the potential to considerably improve the signal-to-background ratio in spectroscopic experiments at the nanoscale.
\end{abstract}


\maketitle

\section{Introduction}
Metallic nanostructures support collective oscillations of the electron gas, which couple strongly to light. At extended metal-dielectric interfaces, the resulting surface-bound modes, termed surface plasmon polaritons (SPPs), may propagate along the interface. At geometric singularities, i.e., in regions of deep-subwavelength radius of curvature, these modes tend to be spatially confined and are called localized surface plasmons (LSPs). During recent years, experimental realizations of SPP guiding in sub-wavelength dimensions~\cite{GramotnevEtAl10} and the transformation of propagating SPPs into LSPs have drawn tremendous attention within the field of plasmonics. The concept of SPP to LSP transformation has been investigated theoretically and experimentally for different metal nanostructures~\cite{AubryEtAl10,Gramotnev05,LuoEtAl10,BabadjanyanEtAl00} in both the adiabatic~\cite{BabadjanyanEtAl00,Stockman04,DurachEtAl07} and non-adiabatic~\cite{IssaEtAl07} limit. Theoretically, it has been shown to be particularly promising in its adiabatic limit, i.e., for wave\-guides with gradually (adiabatically) varying cross-section, for which reflection-free SPP propagation and efficient transformation are predicted~\cite{Stockman04}. This gives rise to a new class of probes in apertureless scanning near-field optical microscopy (SNOM), focusing light down to volumes well below the classical diffraction limit~\cite{BabadjanyanEtAl00,Stockman04}. In particular, it has been predicted that SPP wavepackets may be localized both in space and time when they are launched onto a tapered metallic wave\-guide, e.g., a conical tip with a nanometer-sized apex and sufficiently small opening angle~\cite{Stockman04}. This formation of a strongly confined lightspot is now termed "adiabatic nanofocusing"~\cite{Stockman04}. The expected nanometer scale confinement offers the possibility to efficiently couple light into a single nanoobject, making such probes highly attractive for inherently background-free, new spectroscopic applications on the nanometer scale.

Adiabatic nanofocusing has been demonstrated experimentally by grating coupling of light to SPPs on electrochemically etched gold tapers~\cite{RopersEtAl07}. Such tapers have been incorporated in SNOM setups and nanostructures such as a single gold nanoparticle~\cite{SadiqEtAl11,SchmidtEtAl12} or silicon step edge~\cite{NeacsuEtAl10} have been imaged, demonstrating an optical resolution of down to $10\,\rm nm$. Moreover, time-resolved studies showed that the time structure of ultra-fast laser pulses in the few-femtosecond regime hardly changes upon grating coupling to SPPs and during propagation over distances of several tens of microns on the surface of a gold taper~\cite{SchmidtEtAl12}. These results have been confirmed by three-dimensional finite difference time domain~(FDTD) simulations~\cite{SchmidtEtAl12}. These theoretical investigations and experimental demonstrations suggest that pump-probe studies employing adiabatic nanofocusing are well within experimental reach.

\begin{figure}[b]
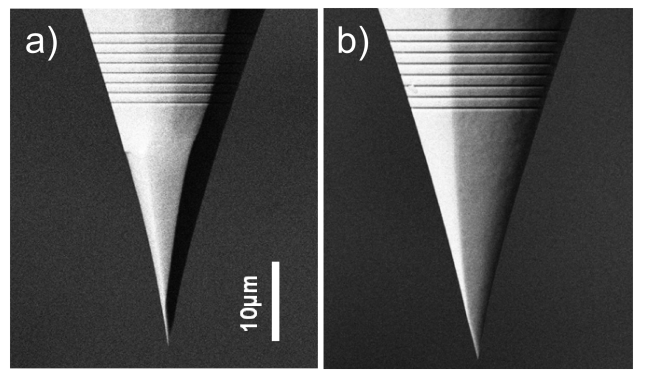

\sglcolfigure{SEMtips}
\caption{Scanning electron micrographs of two gold tapers which were obtained by electrochemical AC-etching of annealed gold wires followed by focused ion beam milling of the grating couplers. Field enhancement at the taper apex was much more pronounced for the tip shown in panel a).}
\label{fig:SEMtips}
\end{figure}

\begin{figure*}[t]
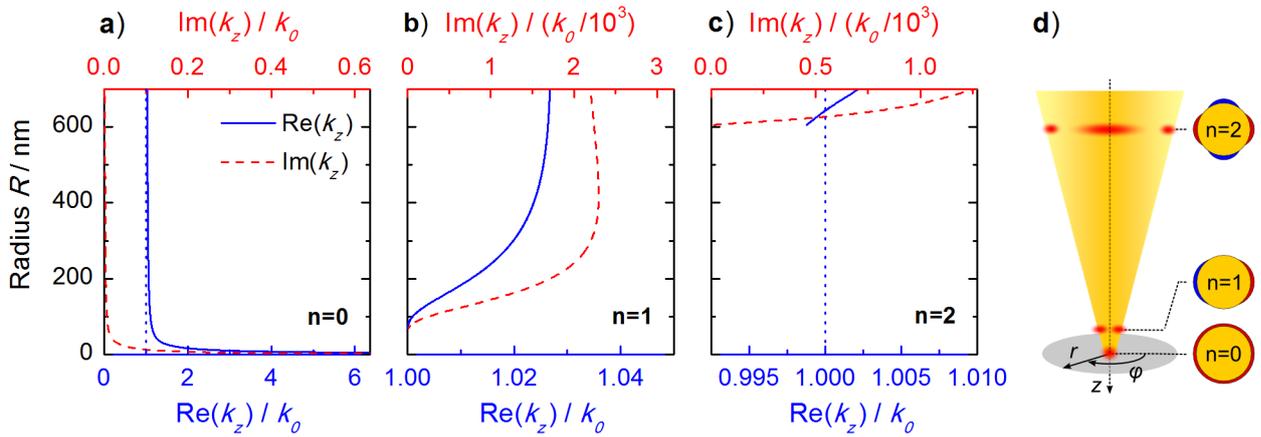

\dblcolfigure{propconstR}
\caption{Propagation constants $k_z$ of the three lowest eigenmodes of an infinitely long gold wire as a function of wire radius $R$. Displayed are the real (solid blue lines) and imaginary (dotted red lines) parts of $k_z$ pointing along the taper axis for angular momenta $n=0,1,2$ (panels a)-c)). The SPP frequency corresponds to a vacuum wavelength of $800\,\rm nm$, as used in the experiment. For $n=0$, both $\mbox{Re}(k_z)$ and $\mbox{Im}(k_z)$ diverge for small radii. For $n=1$, $k_z$ approaches the light line, $k_0=\omega/c_0$, for small radii. A cutoff behavior at approximately $600\,\rm nm$ is observed for $n=2$. Panel d) summarizes these findings and schematically illustrates the mode profiles.}
\label{fig:propconstR}
\end{figure*}

Despite its large potential, adiabatic nanofocusing has not yet been established as a standard approach to optical investigations on the nanoscale. Arguably, the most critical challenge is reproducibility. Taper probes produced by electrochemical etching as described in~\cite{SchmidtEtAl12} differ drastically in their conversion efficiency from far-field radiation to nano-localized LSP intensity at the apex. The differences in conversion efficiency exist even for tips having nominally identical, smooth surfaces, appropriate opening angles and small apex radii. Therefore, experiments are  time consuming and high quality data are scarce. Scanning electron micrographs of two conical taper probes that were used in our own experiments are shown in Figure~\ref{fig:SEMtips}. Whereas their geometry and morphology appear very similar, they showed very different conversion efficiencies.

Conceptually, the difference in conversion efficiency may be explained as a mode matching problem. When employing a grating coupler to excite SPPs, as done, e.g., in~\cite{RopersEtAl07,NeacsuEtAl10,SadiqEtAl11,SchmidtEtAl12}, different eigenmodes of the taper are excited. Their relative amplitudes are given by their spatial overlap with the incoming field and depend critically upon the exact shape of the tip and the incoming phase front. Theoretical studies have shown that only the lowest, rotationally symmetric eigenmode is nanofocused down to the very apex of the tip~\cite{VerhagenEtAl09} where it gets highly confined. In contrast, higher order modes radiate into the far-field while propagating along the taper. This far-field coupling can occur  within the last micron from the tip apex, making it difficult to distinguish these modes from the nano-localized LSP fields. This hampers efficient background-free nanofocusing with the setup described in~\cite{SadiqEtAl11} and higher order mode suppression remains an unsolved challenge.

In this work we present the first steps toward overcoming this challenge. Our approach seeks to provide direct access to the field distribution in the vicinity of the taper apex. To this end, we apply k-space imaging of the fields emitted from the taper apex. By analyzing the rotational symmetry of the resulting radiation patterns we are able to distinguish between contributions from eigenmodes of different orders. In effect, this opens up new possibilities for optimizing the relative contribution from the lowest mode as  well as extracting the near-field contributions at the taper apex. We think that this is an important step toward ultrafast pump-probe spectroscopy on the nanoscale. Specifically, we show that we can differentiate between the zeroth and first order mode, which generally is difficult when using conventional far field imaging techniques.

\begin{figure*}[t]
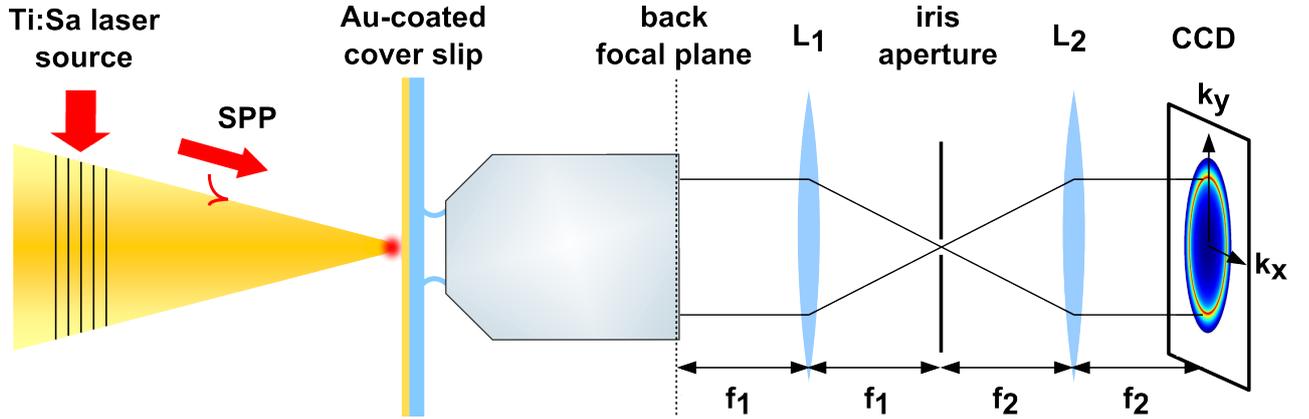

\dblcolfigure{setup-kspace}
\caption{Setup for k-space imaging of radiation emitted from the apex of a sharp gold taper. SPP wavepackets launched via a grating coupler are adiabatically nanofocused. A gold-coated cover slip is brought into close proximity of the taper. Light scattered through the cover slip is collected with a high numerical aperture objective ($\mbox{NA}=1.3$). The spatial Fourier transform of light from the front focal plane is imaged from the back focal plane of the objective onto a CCD camera. A spatial filter is applied in the intermediate real image plane.}
\label{fig:setup}
\end{figure*}

\section{Eigenmodes of a tapered wire}

In order to describe the process of linear SPP propagation on the SNOM probe in the framework of classical electrodynamics, it is useful to expand the SPP wavepacket in terms of orthonormal eigenmodes at any point along the wave\-guide. The concept of adiabatic nanofocusing assumes that the diameter of a tapered wave\-guide varies only slowly over the distance of one wavelength of the guided SPP. Therefore, a proper local set of eigenmodes is well approximated by the eigenmodes of an infinitely long wave\-guide having this local diameter. For a cylindrical metal wire surrounded by a homogeneous dielectric, these solutions are the well-known wire modes~\cite{Stratton41,AshleyEtAl74,PfeifferEtAl74}. These modes possess a discrete rotational symmetry with respect to the angle $\varphi$ of a cylindrical coordinate system whose z-axis coincides with the wire axis. The symmetry is governed by a quantization number $n$, corresponding to an angular momentum which manifests itself in a field distribution of the form $\vec{E}(r,z,\varphi)=\vec{E}(r,z)\cdot\exp(in\varphi)$.
The propagation of the wire modes is described by a transcendental equation which can be derived from Maxwell's equations and the appropriate boundary conditions at the metal-dielectric interface~\cite{Stratton41,AshleyEtAl74,PfeifferEtAl74,VerhagenEtAl09}. It yields the propagation constant $k_z$ along the tapered wire as a function of the dielectric material properties, wire radius and frequency of the SPP mode.

Based on this approach we have numerically calculated the propagation constant for a gradually decreasing wire radius $R$. The results for an SPP frequency corresponding to a vacuum wavelength of $800\,\rm nm$ on a gold wire (dielectric constant $\epsilon_1=-24.7470 + 1.8834i$, obtained by fitting experimental data from~\cite{JohnsonEtAl72}) surrounded by air ($\epsilon_2=1$) are displayed in Figure~\ref{fig:propconstR}. 

For decreasing radius, the three lowest modes $n=0,1,2$ (cf. panels a-c) behave distinctly differently. For the lowest, rotationally symmetric mode with $n=0$, both the real and imaginary part of the propagation constant are divergent for vanishing wire radius. Thus, this mode remains strongly bound to the interface while its wavelength shortens, leading to a decrease in SPP group velocity. In effect, the SPP is transformed into a strongly confined LSP. In the region below the tip, the field emitted by this LSP mode closely resembles that of an oscillating point dipole~\cite{KoglinEtAl97,EstebanEtAl06}. In the case of a tapered wave\-guide, this dipole moment is oriented along the taper axis. Hence, the field as well as the radiation pattern emerging from the apex maintain the rotational symmetry of the eigenmode.

In contrast, for the $n=1$ mode, the propagation constant approaches the light line, $k_0=\omega/c_0$, for radii of about $70\,\rm nm$, indicating that the mode looses its bound character. It is not transformed into a highly confined LSP at the taper end and we therefore refer to it as a loosely bound photonic mode.
The third mode, with $n=2$, displays a cut-off behavior already for a radius slightly above $600\,\rm nm$. This mode is not sustained on the wire for smaller radii. The same is true for all higher order modes.

As a consequence, only the $n=0$ and $n=1$ eigenmodes of the wire are expected to contribute to near-fields in the vicinity of the taper apex. When scattered into the far-field, the contributions of those two modes cannot be distinguished by only using far-field optics as their spatial separation is well below the classical diffraction limit. When using tapered nanofocusing probes in a SNOM setup, just the $n=0$ eigenmode generates the desired, spatially highly resolved signal, containing information about the sample properties on the nanoscale. The $n=1$ mode, however, is expected to contribute to a non-negligible background signal due to its weak confinement.

\section{Experimental setup}

Our experimental setup used for k-space imaging is sketched in Figure~\ref{fig:setup}. The field distribution at the taper apex is probed by placing it in front of a gold-coated dielectric surface, which couples some fraction of the evanescent near-fields at the apex to propagating far-field modes. The angular distribution of these propagating waves is measured with an immersion objective. In the following, we describe the individual parts of the experimental setup.

\begin{figure*}
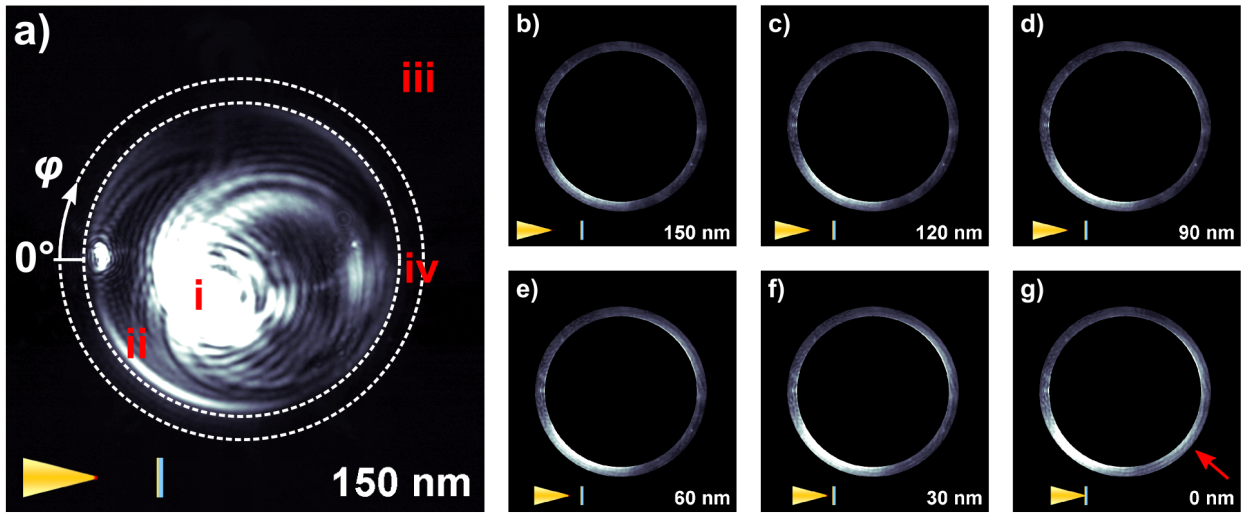

\dblcolfigure{sequence}
\caption{k-space patterns recorded while approaching a gold-coated cover slip to the SNOM probe. Tip-sample spacings are indicated in the lower right corner of the respective images and the corresponding pictographs in the lower left corner. Panel a) shows raw data, the dashed circles mark the region of interest (ROI, labeled iv). Panels b)-g) represent the evolution of the k-space pattern during approach. For clarity, the data are limited to the ROI and are rescaled to the 90th percentile of all pixel intensities within the ROI in panel g).}
\label{fig:frames}
\end{figure*}

Tapers as shown in Figure~\ref{fig:SEMtips} are produced from single-crystalline gold wire by an electrochemical AC-etching technique followed by focused ion beam milling of a grating coupler~\cite{SchmidtEtAl12}. Typical cone opening angles are between $20^\circ-30^\circ$ and apex radii are well below $30\,\rm nm$. These tips have a very smooth surface and accordingly high SPP propagation lengths as reported earlier~\cite{SchmidtEtAl12}.

To excite SPPs on the probe, we use a pulsed titanium sapphire (Ti:Sa) laser source operating at a central wavelength of $800\,\rm nm$ and featuring a $10\,\rm nm$-wide Gaussian spectral distribution. This light is focused onto the grating coupler with a microscope objective ($\mbox{NA}=0.1$), applying an average power of $200\,\mu\rm W$ at a repetition rate of $80\,\rm MHz$. The SPP wavepacket launched on the coupler propagates toward the taper apex.

The light emitted from the taper apex is then transmitted through a $20\rm nm$ thick gold film on a cover slip. The main reason for using a Au-coated sample was to enhance the evanescent light contributions to the k-space images. The gold film has three major effects on those images.  (i) It substantially suppresses the transmission of propagating waves ( $k_\parallel < k_0$ , allowed light) which have to tunnel through the film. For our film, we measure that the transmission is reduced to about 20\%. (ii)  It enhances the total power radiated into forbidden light modes by about a factor of 2, as can be seen from a simple estimate based on a point dipole model (s., e.g.,\cite{NovotnyEtAl06}, Ch. 10). (iii) Most importantly, it allows us to probe the near-field coupling between the tip dipole, i.e., the near fields at the tip apex, and its image dipole induced in the metal (s.~\cite{KnollEtAl00,RaschkeEtAl03}).

The light transmitted through the gold-coated cover slip is collected using an oil immersion objective with high numerical aperture ($\mbox{NA}=1.3$, Olympus UPlanFLN 100x). The tip apex position lies within the focal plane of the microscope objective. In this configuration, the objective performs a Fourier transform of the field distribution from the front into the back focal plane. Two additional lenses with focal lengths of $f_1=150\,\rm mm$ and $f_2=100\,\rm mm$, respectively, image the back focal plane onto a CCD camera (Thorlabs DCC1545M-GL) in a 4f-configuration. In the intermediate real image plane, filtering of scattered light from the grating coupler is realized with an iris aperture.

The distance between probe and cover slip is controlled using a tuning fork-based force sensor in a non-contact-mode atomic force microscope (AFM). This microscope is a modified version of the setup described in~\cite{SadiqEtAl11}. The taper probe is attached to one prong of a quartz tuning fork that oscillates with a peak-to-peak amplitude of $1\,\rm nm$. The cover slip is mounted onto a three-axis piezoelectric stage (PI P-363.3CD). This allows slowly approaching the sample to the taper over a distance of several hundreds of nanometers in steps of $30\,\rm pm$ until the tuning fork starts to be damped by tip-sample interactions. The damping occurs on a length scale of less than $5\,\rm nm$. The sample is subsequently retracted to its original position with the same step size. While recording force-approach curves, the CCD camera is triggered to capture the respective k-space images for selected positions on the approach curve.

\begin{figure*}
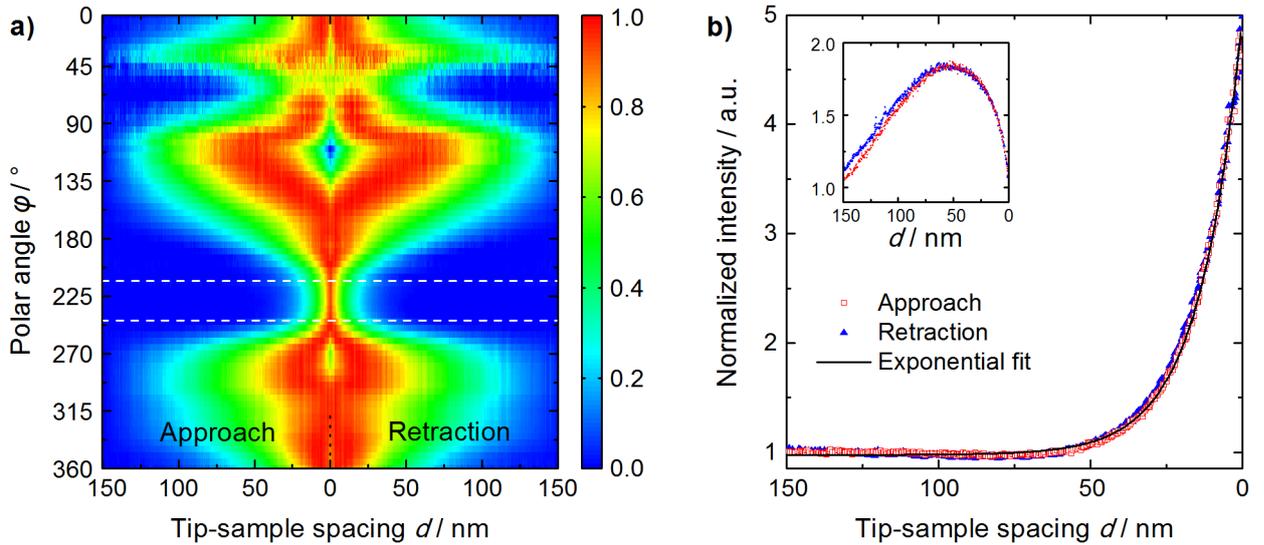

\dblcolfigure{DecayGraphsWithInset3}
\caption{Quantitative analysis of the evolution of k-space patterns during tip approach and retraction. Panel a) shows the angular intensity distribution, radially integrated across the ROI (cf. Figure~\ref{fig:frames}a), as a function of tip-sample spacing. At each angle, the data have been normalized. The region between the two dashed white lines has been averaged with respect to the angle and the data have been plotted in panel b) together with an exponential fit to the part of the data recorded during approach. The decay length is $14\,\rm nm$. Here, the data have been normalized to the background signal. The inset shows an equal set of data for a different angular region around $110^\circ$.}
\label{fig:decayanalysis}
\end{figure*}

\section{Experimental Results}

We have recorded distance-dependent k-space images of the light transmitted through the gold-coated cover slip for several grating-coupler type nanofocusing tapers. The results of a representative experiment for the taper shown in Figure~\ref{fig:SEMtips}a) are described in detail in the following. A sequence of camera images was recorded at selected positions on the AFM approach curve. The first frame of this sequence taken at a tip-sample spacing of about $150\,\rm nm$ is displayed in Figure~\ref{fig:frames}a). In this image, the bright pattern (labeled position i) corresponds to radiation from the far-field with in-plane k-vectors~$k_\parallel$ smaller than $k_0=\omega/c_0$. The ring-like pattern (labeled ii) is associated with in-plane k-vectors~$k_\parallel$ slightly larger than $k_0$. It can be attributed to leakage radiation from SPPs excited on the gold film~\cite{SimonEtAl76,HechtEtAl96,DrezetEtAl06}. The dark area outside the larger dashed circle (labeled iii) reflects in-plane k-vectors~$k_\parallel$ larger than $1.3\cdot k_0$, which are not captured by our oil immersion objective. The area between the two dashed circles (labeled iv) is thus the region of interest (ROI) to us. Here, only forbidden light~\cite{HechtEtAl96} originating from evanescent fields at the taper apex is detected.
Panels b)-g) of Figure~\ref{fig:frames} show a sequence of six camera images with equidistant spacing along the corresponding AFM approach curve. For clarity, they exclusively depict the ROI and the intensities have been rescaled. The red arrow in panel g) marks a short-ranged increase in intensity which is only visible in this panel.

A quantitative analysis of the full sequence of images is displayed in Figure~\ref{fig:decayanalysis}. It has been obtained by radially integrating across the ROI for each position on the approach curve. The resulting integrated intensities are plotted as a function of in-plane angle $\varphi$ and tip-sample spacing $d$. To improve comparability and to simplify identification of the individual features, each horizontal line, corresponding to a fixed angle $\varphi$, was normalized to its maximum after subtraction of its first value during approach (at $d=150\,\rm nm$). The color bar has been restricted to the range from zero to one.

Similar plots have been generated from measurements at different spatial positions on our gold film. Within our signal-to-noise ratio, the data show no variation with sample position. This suggests that the possible excitation of localized surface plasmon fields on the gold film has only a very limited effect on our experimental data and supports that our experiments indeed map the emission properties of the nanotaper.

Two qualitatively different decay lengths emerge from these plots. One of them is well below $30\,\rm nm$ and appears most prominently at polar angles around $225^\circ$. This area is marked by the red arrow in Figure~\ref{fig:frames}g). The other, much longer decay length occurs in two opposing lobes above and below the region marked by the two dashed white lines. These two different decay lengths have qualitatively also been observed in experiments performed with other tapers. To quantify the short decay length for the given tip, the data inside the marked zone in Figure~\ref{fig:decayanalysis}a) have been azimuthally  integrated. For comparison this has also been done with the data between $100^\circ$ and $120^\circ$. The integrated values are plotted in panel b) and the inset therein, respectively. In this panel, the offset at $d=150\,\rm nm$ was not subtracted and was used for normalization to it such that the y-axis gives the relative increase with respect to this offset. An exponential fit to the data points taken during approach (open red squares), offset by $1.0$, has been added. The corresponding decay length is $14\,\rm nm$ and the overall increase in intensity amounts to a factor of five.

\section{Discussion}

The angular decay spectrum displayed in Figure~\ref{fig:decayanalysis}a) strongly suggests that only those SPP fields that are coupled to the lowest order and first higher order eigenmode of the taper propagate as bound modes into close proximity of the taper apex. This is qualitatively understood in the following way: 
The symmetry of the $n=1$ mode is that of an in-plane dipole with its dipole moment oriented parallel to the sample surface. Thus, its angular radiation distribution should exhibit two opposing lobes. These lobes are clearly seen in Figure~\ref{fig:decayanalysis}a) in the quadrants ranging from $90^\circ -180^\circ$ and $270^\circ -360^\circ$, respectively. The decay length observed in these two lobes agrees with the expectation for the evanescent decay in air related to the in-plane k-vector components $k_\parallel$ within the ROI.

For the eigenmode with $n=0$ a rotationally symmetric radiation pattern is expected for an ideally symmetric tip. Theoretically, it has been shown that the fields at the tip apex decay with a length slightly shorter than the apex radius~\cite{EstebanEtAl07,BehrEtAl08,ParkEtAl13}. Taking image charge effects into account~\cite{KnollEtAl00,RaschkeEtAl03}, the decay in k-space closely resembles the behavior in real space. This short-ranged decay is most prominently encountered in the region marked with the dashed white lines in Figure~\ref{fig:decayanalysis}a) and the red arrow in Figure~\ref{fig:frames}g). In this region, the line shape nicely follows an exponential decay as can be seen in Figure~\ref{fig:decayanalysis}b). The decay length of $14\,\rm nm$ is in convincing agreement with the theoretical expectation. As determined by scanning electron microscopy, our tapers usually have apex radii between $10$ and $20\rm nm$. For the region between $0^\circ$ and $50^\circ$, the measured total intensities are small. Therefore, the values obtained in this region should be considered with care.

Interference between the two modes with $n=0$ and $n=1$ is expected to give rise to constructive interference in one lobe and destructive interference in the other, resulting in an asymmetric distribution of the intensity in the two lobes as well. In Figure~\ref{fig:frames}, an asymmetry in the intensity of the two opposing lobes is indeed observed. By comparison with the overall intensity detected from the $n=0$ mode, however, it can be deduced that this effect still cannot fully account for the observed asymmetry.

As can be seen in Figure~\ref{fig:frames}a), an asymmetric signature of surface plasmon polariton resonance leakage is also detected in the raw data (labeled ii). The tails of this resonance extend into that part of k-space which we define as the region of interest. A substantial amount of the asymmetry in the processed signal (visible in Figure~\ref{fig:frames}b)-g)) can be attributed to this effect. The excitation of this resonance is most likely due to scattering of radiation from the grating coupler as well as emission from higher order modes of the taper. It is therefore possible that this spurious signal overlaps in our k-space images with the desired signatures from the $n=0$ and $n=1$ modes and thus contributes to the recorded asymmetry.

To give an estimate of the relative contributions of the $n=0$ and $n=1$ mode, we compare the intensities in our k-space images. The relative amplitude of the $n=0$ and $n=1$ mode emission appears similar when approaching the surface. These k-space images probe evanescent fields within a finite wave vector range, extending to about $1.3\cdot k_0$. The k-space distribution of the (spatially highly confined) $n=0$ mode extends to much larger k-vectors than that of the $n=1$ mode. This k-spectrum is much broader than what is probed in our experiments. Hence only a fraction of the $n=0$ mode emission can be detected. In high-resolution near-field experiments, however, the entire k-spectrum contributes.

Inside the two lobes with the long-ranged decay, a decrease of the intensity sets in for small values of the tip-sample spacing (s. inset in Figure~\ref{fig:decayanalysis}b)). To account for this behavior, we believe that we need a near-field source producing a field with a k-space symmetry similar to that of the $n=1$ mode. Its field amplitude should vary substantially over the last $20\rm nm$. Such fields would be generated by a laterally oriented dipolar emitter located near the tip apex. This dipole emission may reflect contributions of the $n=1$ mode which are weakly guided towards the taper apex. 
Earlier FDTD simulations~\cite{LeeEtAl11} indeed indicate that, for asymmetric grating excitation on one side of the taper, as done in our experiments, the field near the apex is not just that of a point-like dipole purely oriented along the z-axis. Instead, a slightly asymmetric near-field near the tip apex is predicted. The formation of a laterally oriented dipole field is clearly observed in these simulations. We believe taking these effects into account can readily explain the interference effects observed in the two lobes in Figure~\ref{fig:decayanalysis}a).

In view of the discussion above, the results presented in Figure~\ref{fig:decayanalysis}a) are well understood considering the emission from only the lowest two eigenmodes of the conical taper. Apparently, all higher order modes decouple from the taper already at distances of the order of one wavelength or above. Hence, they contribute only weakly to the recorded forbidden light images presented here. This suggests that k-space filtering is an efficient means for suppressing undesired background fields in adiabatic nanofocusing SNOM.

\section{Conclusion}

In summary, we have presented a method to investigate the radiation patterns of adiabatic nanofocusing SNOM probes. Our experiments demonstrate that SPP fields coupled to the lowest two eigenmodes are efficiently funneled toward the taper apex. Most importantly, we show that the $n=0$ eigenmode produces a spatially highly localized near-field at the apex, as predicted by theory, whereas the loosely bound $n=1$ mode is much less confined. Their contributions are readily identified by analyzing the evolution of angular radiation patterns over the course of an approach curve. This opens up the possibility to separate the rapidly decaying $n=0$ mode from the undesired higher modes by k-space filtering.

These findings form the basis for further improvements and novel applications. The rapidly decaying contribution to the k-space images provides a direct measure of the desired $n=0$ mode. Hence, we can use this signal to optimize adiabatic nanofocusing into the lowest order mode. This minimizes background from higher order modes and thus provides a major advancement towards background-free scanning near-field optical spectroscopy. Since we are able to further separate the near-field contributions by k-space filtering, we can now apply this method to spectroscopic studies of the coupling between nano-localized fields and single quantum objects. Such experiments are currently underway in our laboratory.

\begin{acknowledgments}
Financial support by the Deutsche Forschungsgemeinschaft (SPP1391 `Ultrafast Nanooptics', DFG-Li860/5-1 and DFG-NSF Materials World Network), by the Korea Foundation for International Cooperation of Science and Technology (Global Research Laboratory Project K20815000003) and by the European Community through the CRONOS project, Grant Agreement
No. 280879, is gratefully acknowledged. ME acknowledges funding from the Studienstiftung des Deutschen Volkes.\\
We thank Heiko Kollmann for supporting the theoretical calculations, Kerstin Reiners and Marco Ohm for contributions at an early phase of this work and Raimond Angermann for continuous high quality technical support. 
\end{acknowledgments}

\bibliography{BeilsteinSubm13_V2}
\vspace{3cm}

\end{document}